\documentclass[11pt]{article}
 \pdfoutput=1
\usepackage{jheppub}

\usepackage{amsmath}
\usepackage{verbatim}   
\usepackage{subfigure}  

\usepackage{amsfonts}
\usepackage{amssymb}
\usepackage{mathrsfs}
\usepackage{graphicx}
 \usepackage{slashed}
 \usepackage{epsfig}
 \usepackage{url}

\newcommand{\newc}{\newcommand}
\newc{\gsim}{\lower.7ex\hbox{$\;\stackrel{\textstyle>}{\sim}\;$}}
\newc{\lsim}{\lower.7ex\hbox{$\;\stackrel{\textstyle<}{\sim}\;$}}
\newc{\gev}{\,{\rm GeV}}
\newc{\mev}{\,{\rm MeV}}
\newc{\ev}{\,{\rm eV}}
\newc{\kev}{\,{\rm keV}}
\newc{\tev}{\,{\rm TeV}}

\def\tr{\mathop{\rm tr}}

\newc{\mz}{M_Z}
\newc{\mpl}{M_*}
\newc{\mw}{m_{\rm weak}}
\newc{\nr}[1]{N^c_R{}_{#1}}
\usepackage{amsmath}

\def\beq{\begin{equation}}
\def\eeq{\end{equation}}
\def\bea{\begin{eqnarray}}
\def\eea{\end{eqnarray}}
\def\bitem{\begin{itemize}}
\def\eitem{\end{itemize}}
\newcommand{\bec}{\begin{center}}
\newcommand{\eec}{\end{center}}
\newcommand{\kahler}{K\"{a}hler~}

\newcommand{\half}{\frac{1}{2}}

  \newcommand{\GeV}{{\mathrm {GeV}}}

\def\bar#1{\overline{#1}}
\def\vev#1{\left\langle #1 \right\rangle}

\def\inv{^{\raise.15ex\hbox{${\scriptscriptstyle -}$}\kern-.05em 1}}
\def\lbar{{\lower.35ex\hbox{$\mathchar'26$}\mkern-10mu\lambda}} 

\def\to{\rightarrow}

\let\<=\langle
\let\>=\rangle

\let\+=\uparrow

\let\la=\lambda

\newcommand{\Lag}{{\mathcal L}}

\def\slashchar#1{\ensuremath{                               %
   \setbox0=\hbox{${}#1{}$}       
   \dimen0=\wd0                                 
   \setbox1=\hbox{/} \dimen1=\wd1               
   \ifdim\dimen0>\dimen1                        
      \rlap{\hbox to \dimen0{\hfil/\hfil}}      
      {}#1{}                                    
   \else                                        
      \rlap{\hbox to \dimen1{\hfil${}#1{}$\hfil}}   
   \fi}}

\begin{document}
\hfill \vspace{-5mm} SITP-12/43

\title{Axion Mediation}

\author[a]{Masha Baryakhtar,}
\emailAdd{mbaryakh@stanford.edu}
\author[b]{Edward Hardy,}
\emailAdd{e.hardy12@physics.ox.ac.uk}
\author[a,b]{John March-Russell}
\emailAdd{jmr@thphys.ox.ac.uk}
\affiliation[a]{Stanford Institute for Theoretical Physics, Department of Physics,\\
 Stanford University, Stanford, CA 94305, USA}
\affiliation[b]{Rudolf Peierls Centre for Theoretical Physics,
University of Oxford,\\
1 Keble Road, Oxford,
OX1 3NP, UK}

\date{\today}

\abstract{ We explore the possibility that supersymmetry breaking is
  mediated to the Standard Model sector through the interactions of a
  generalized axion multiplet that gains a F-term expectation
  value. Using an effective field theory framework we enumerate the
  most general possible set of axion couplings and compute the
  Standard Model sector soft-supersymmetry-breaking terms.  Unusual,
  non-minimal spectra, such as those of both natural and split
  supersymmetry are easily implemented. We discuss example models and
  low-energy spectra, as well as implications of the particularly
  minimal case of mediation via the QCD axion multiplet. We argue that
  if the Peccei-Quinn solution to the strong-CP problem is realized in
  string theory then such axion-mediation is generic, while in a field
  theory model it is a natural possibility in both DFSZ- and KSVZ-like
  regimes.  Axion mediation can parametrically dominate
  gravity-mediation and is also cosmologically beneficial as the
  constraints arising from axino and gravitino overproduction are
  reduced. Finally, in the string context, axion mediation provides a
  motivated mechanism where the UV completion naturally ameliorates
  the supersymmetric flavor problem.
    
  }


\maketitle

\section{Introduction}
\label{sec:intro}
%

If supersymmetry (SUSY) is realized at the TeV scale and is wholly or
in part responsible for the solution to the hierarchy problem then,
given experimental constraints, its low-energy realization is almost
certainly more complex than has typically been assumed in much of the
literature.  For instance, the constrained minimal supersymmetric
standard model (CMSSM), even with altered assumptions in the Higgs
sector, seems unlikely to be close to the truth.  If one closes one's
eyes to the question of a consistent and motivated UV realization, a
number of low-energy spectra can be constructed which lead to a
sufficient weakening of the LHC search limits so that SUSY is still
relevant for the hierarchy problem.  Assuming for definiteness that
R-parity is conserved in the low-energy theory, so-called ``natural",
``compressed" and ``stretched" SUSY spectra have been argued to
significantly reduce the LHC limits while maintaining relatively low
fine-tuning
\cite{Dimopoulos:1995mi,Cohen:1996vb,Martin:2007gf,Lodone:2012kp,Hall:2012zp}
(possibly with a modified NMSSM-like Higgs sector
\cite{Ellwanger:2009dp}, either in the large $\lambda$ limit
\cite{Hall:2011aa,Barbieri:2006bg,Barbieri:2007tu,Hardy:2012ef}, or
with an altered R-symmetry structure
\cite{Kribs:2007ac}).\footnote{Relaxing the assumption of R-parity
  conservation weakens LHC limits only by a modest amount, with
  gluinos often still heavier than a TeV; it may even increase the
  limits in the case of leptonic R-parity violation
  \cite{ref:cmsrpv,Barbier:2004ez}.  Significantly reducing the LHC
  limits requires SUSY events with nearly no missing energy and few
  leptons, both of which appear in generic spectra from decays of
  charginos and tops.}
  
However, the best-known and most studied mediation mechanisms for
supersymmetry breaking, namely gauge mediation, anomaly mediation, and
minimal sequestered gravity mediation, while consistent with the
severe indirect flavor constraints on supersymmetry breaking (by
construction in the gravity case) have trouble realizing the modified
spectra.  It is therefore worthwhile to look further for motivated
mediation schemes which lead to one or more of the variant spectra.
In this paper we argue that an axion multiplet, possibly the QCD axion
multiplet itself, is a natural candidate to mediate supersymmetry
breaking and can lead to unusual and phenomenologically attractive
patterns of soft terms.

The QCD axion provides the best known solution to the strong-CP
problem and so is a very well-motivated extension to the Standard
Model (SM) \cite{Peccei:1977hh,Weinberg:1977ma,Wilczek:1977pj}.  The
axion mechanism requires a new scale $f \ll M_{\rm{Pl}}$ at which
Peccei-Quinn (PQ) symmetry is spontaneously broken, and results in a
new light particle which interacts with the SM.  In the supersymmetric
extension of such models the axion multiplet can mediate supersymmetry
breaking to the visible sector and the separation of scales of the
axion and $ M_{\rm{Pl}}$ ensures that axion mediation will dominate
over uncontrollable gravity-mediated effects.  Moreover, the axion
sector itself provides a natural candidate for the origin of
supersymmetry breaking.  Thus, the QCD axion is a promising yet
relatively unexplored candidate to dominate supersymmetry
breaking.\footnote{Models in which the axion couples to the SUSY
  breaking sector, and the messengers both mediate SUSY breaking
  and generate the anomalous QCD coupling were studied in
  \cite{Higaki:2011bz}.}

The structure of the paper is as follows: in the remainder of the
Introduction, we discuss how a UV theory typically results in one or
more axion multiplets, and mechanisms by which axions may gain
significant F-terms.  In Section~\ref{sec:couplings}, we set our
notation and define the effective interactions of an axion multiplet
consistent with shift symmetry, paying particular attention to
invariance under local chiral field redefinitions.  Following this
Section~\ref{sec:F-term} gives expressions for the soft terms
generated in the MSSM sector when a generic axion multiplet gains the
dominant F-term, including typical hierarchies of sfermion and
gaugino mass terms. In Section~\ref{sec:spectra} we specialize the
generic couplings to some well motivated cases which lead to
non-standard spectra of soft terms, and briefly discuss the resulting
phenomenology.  Finally, in Section~\ref{sec:goldstino} we show there
are some cosmological advantages to such a set-up, and conclude in
Section~\ref{sec:conclusion}.

\subsection{Motivation and UV Completions}
\label{sec:uv}

In light of the strong CP problem and the necessity of a QCD axion, we
note that there are a variety of models that can generate axions with
couplings to SM fields. In particular, we consider both string models
in which axion multiplets are the moduli multiplets of
the underlying compactification, and purely field-theoretic
models. For both possibilities we discuss mechanisms by which the
axion multiplet may acquire a large F-term and hence dominate the
mediation of supersymmetry breaking to the visible sector.
 
In the field theory case an anomalous global U(1) symmetry---the PQ
symmetry---is broken and an axion arises as a pseudo-goldstone
boson. In KSVZ models the axion sector is not coupled to the visible
sector chiral multiplets, and the anomalous coupling between the axion
multiplet and the standard model gauge groups is a result of
integrating out new heavy vector-like pairs of chiral multiplets
charged under the PQ symmetry and the Standard Model gauge groups
\cite{Kim:1979if,Shifman:1979if}. In DFSZ models the MSSM matter
itself is charged under the PQ symmetry and generates the anomalous
coupling \cite{Dine:1981rt,Zhitnitsky:1980tq}. A realistic UV theory
may be expected to be some mixture of these two cases. Traditionally
QCD axion models have been considered in the `axion window',
$10^{9}\gev \lsim f \lsim 10^{12}\gev$: the parameter regime in which
both experimental search limits are satisfied (the lower bound) and
axions are not more than the observed amount of dark matter for
generic values of the axion field in the early universe (the upper
bound).

In field theoretic models the axion will mediate 
supersymmetry breaking if a single sector spontaneously breaks a
global symmetry leading to a Nambu-Goldstone boson (the axion) and
simultaneously breaks supersymmetry such that the axion gains an
F-term, $F_A$.  It is quite plausible that one sector can accommodate
both of these roles in global supersymmetry, as any sector that
breaks a global symmetry and has no run-away directions will
necessarily also break supersymmetry so that the Nambu-Goldstone boson
can remain massless while the moduli direction gains a mass.  This, in
fact, is used as a criterion in the search for theories that break
supersymmetry in strongly coupled regimes (for a review of dynamical
supersymmetry breaking see, for example, \cite{Poppitz:1998vd} and the
references therein).

Moreover, our requirement that the interactions of the axion dominate
the mediation of supersymmetry breaking to the SM sector can be
naturally accommodated in this scenario.  For example, in an
extra-dimensional context, if the supersymmetry breaking sector is
physically sequestered from the SM sector, the multiplets that remain
light are the dominant source of mediation. This fact is due to the
exponential mass suppression of wave functions along the extra
dimension, or in 4D language, the statement that heavy fields gain
large anomalous dimensions and therefore have suppressed interactions
with SM sector. The axion multiplet, protected by Goldstone's theorem,
is precisely such a light field: the axion remains essentially
massless and the other fields in the multiplet only gain masses of
order $\sqrt{F_A}$, while other multiplets can gain masses of order the
axion decay constant $f$ and for $\sqrt{F_A} \ll f$ ($i.e.$ MSSM
soft terms at the TeV scale) the axion dominates the mediation.  A
model of supersymmetry breaking where the low energy degrees of
freedom include an axion has been studied for example in
\cite{Dudas:1993mm}.

In the string context, the topological complexity of a typical
compactification gives rise to many axion-like multiplets.  Each
non-trivial cycle of the internal manifold can lead to a state with
axion-like couplings from the zero mode of an asymmetric tensor field
integrated on the cycle.  The number of such cycles is typically
$\mathcal{O}\left(10^3\right)$ or as large as $\mathcal{O}\left(10^5
\right)$, giving in principle a comparable number of axion-type
fields \cite{Svrcek:2006yi,Arvanitaki:2009fg}.  For instance, in type
IIB string theory, the integral of a rank-4 antisymmetric tensor field
$C_4$ over every independent 4-cycle $\Sigma^4_i$ of the internal
manifold gives rise to an independent pseudo-scalar field $a_i$ in
4-dimensions,
\begin{equation}
\label{eq:axiondef}
a_i = {1\over 2\pi} \int_{\Sigma^4_i} C_4 .
\end{equation}
In addition, the underlying Abelian gauge symmetry in 10D, $C_4\to C_4
+d \lambda_3$, implies that each field $a_i$ inherits a global `PQ'
shift symmetry.  Generally speaking, for a 10D rank-$n$ antisymmetric
field, every independent $n$-cycle of the compactification gives rise
to an independent pseudoscalar mode in 4D. Therefore, a
plethora of 4D axion-like fields results.
 
Not all these potential axion-like modes survive to the low-energy
theory.  As we discuss below, depending on the process of
stabilization moduli can be projected out of the light spectrum.
Typically, however, many remain massless at the perturbative level,
acquiring a (possibly ultra-light) mass due only to non-perturbative
effects which violate the shift symmetry.  Therefore a multitude of
axions can in general survive in the low-energy theory.\footnote{ Such
  axiverse scenaria are discussed in a variety of UV contexts
  \cite{Cicoli:2012sz,Acharya:2010zx,Higaki:2011me}.}
 
The ability of the axion to solve the strong-CP problem, i.e. set
$\bar{\theta}_{CP}$ to less than $10^{-9}$, requires that
non-perturbative effects which explicitly break PQ symmetry are
dominated by QCD dynamics and not other sources such as string
instantons.  Numerically, this suppression of string instantons
translates to the requirement $S\gsim 200$ where $\Lambda^4 = \mu^4
e^{-S}$ with $\Lambda$ the scale that appears in the axion potential
and $\mu$ the UV energy scale, in this case the string scale.  As
argued by Svrcek and Witten \cite{Svrcek:2006yi} one then quite
generally finds that the axion scale $f$ is polynomially suppressed $f
\sim {M_{\rm pl}}/{S}$.  Therefore the QCD axion in string theory may
be expected to have $f\lesssim 10^{16}\gev$.  Even if the QCD axion is
not responsible for mediation of SUSY breaking, once the existence of
a suitable QCD axion is imposed on the string compactification, there
is naturally a plentitude of light axion fields---the string axiverse
---with $f \sim {M_{\rm pl}}/{S}$ \cite{Arvanitaki:2009fg}, and
one or more of these can take part in SUSY-breaking dynamics and
transmit this breaking to the visible sector.

Obtaining string axions in the axion window requires warped
Randall-Sundrum-like throats in which the axion degree of freedom is
IR localized \cite{Flacke:2006ad,Svrcek:2006yi} and whose throat
parameters generate an IR scale $ f \lsim 10^{12}\gev$.  Although such
throats are not unlikely given our limited knowledge of realistic
compactifications \cite{Hebecker:2006bn}, it is another requirement
that one must impose.  As emphasized
in~\cite{Tegmark:2005dy,Hertzberg:2008wr, Arvanitaki:2010sy,Banks:2002sd}, however,
the upper bound of the axion window is sensitive to untested
assumptions about early universe dynamics, and in particular the
over-closure limit is easily relaxed by a mild environmentally-selected
tuning of the post-inflationary mis-alignment angle of the axion
field.  We thus consider that axion scales in the string-motivated
range, $f\sim 10^{16}\gev$, are allowed and are particularly
interesting to explore (for search possibilities in this regime see,
e.g., \cite{Arvanitaki:2009fg,Arvanitaki:2010sy,Graham:2011qk}).

String axions may naturally gain an F-term in the process of moduli
stabilisation: as we now briefly review, this is the classic scenario
of soft terms from moduli fields in string compactifications (see, for
example
\cite{Brignole:1995fb,Brignole1994125,Choi:2005ge,Conlon:2005ki,Ibanez:1992hc,deCarlos:1992da,Kaplunovsky:1993rd}).
In heterotic models this process can occur through gaugino
condensation: a hidden gauge group coupling runs strong at a scale
$\Lambda$ below the string scale, $\Lambda\ll m_S$.  While gaugino
condensation itself does not break supersymmetry, the couplings to the
axion field in the gauge kinetic function can induce an F-term
\cite{Dine198555}. There is a well-known issue of stabilizing the
resulting run-away moduli direction towards infinite field values
\cite{Banks:1994sg}, which may be addressed, for example, through
multiple gaugino condensates.  It has been argued that M-theory
compactifications may lead to an axion multiplet gaining the dominant
F-term \cite{Acharya:2006ia,Acharya:2007rc,Acharya:2008bk}.
Alternatively, flux compactifications of IIB string theories can lead
to moduli with F-terms, either in the process of stabilization as in
large volume scenarios \cite{Balasubramanian:2005zx}, or as a result
of uplifting from a SUSY preserving anti-de-Sitter vacuum to a
Minkowski vacuum, for example in KKLT compactifications
\cite{Kachru:2003aw} (a general review can be found in
\cite{Douglas:2006es}).

Despite much work studying string theory compactifications, there is
still considerable uncertainty in how realistic low energy behavior is
realized. However, since the axion decay constant $f$ must be below
$M_{\rm pl}$, the moduli multiplets that form axions can naturally
dominate the mediation. Additionally, if the QCD axion originates from
string theory, it is very natural that such a multiplet gains a large
F-term: it was shown in \cite{Conlon:2006tq} that if a string axion is
to be the QCD axion, the associated modulus field must be stabilized
non-supersymmetrically. On the other hand, if stabilization occurs
supersymmetrically (though difficult to realize in explicit models),
the QCD axion must come from a field theory sector which, as
discussed, is also a good candidate to dominate supersymmetry breaking
in the MSSM.

A final attractive scenario that can apply to either string or field
theory constructions is if the sector that generates the axion also
strongly stabilises the flat direction of the modulus but does not
generate the dominant supersymmetry breaking of the theory. This can
occur with no SUSY breaking since supergravity in AdS allows
supersymmetric mass splittings within multiplets (before
uplifting). Now if, in addition to this sector, the axion couples to a
sector which experiences gaugino condensation at a scale $\Lambda_h\ll
f$ such that
\begin{equation}
  \label{eq:8}
  \left|\vev{\bar{\lambda}_h\lambda_h}\right|\sim \Lambda_{h}^3,
\end{equation}
then a standard anomalous coupling $ \int d^2 \theta \frac{A}{f}
W_{\alpha}W^{\alpha}$ (directly present or induced through fermion
anomaly diagrams) will lead to an axion F-term
\begin{equation}
  \label{eq:9}
  F_A\sim \frac{\vev{\bar{\lambda}_h\lambda_h}}{f}.
\end{equation}
Unlike the case of heterotic compactifications, there is no danger of
a run-away direction since the moduli direction is stabilized in the
axion sector.  In this model the axion multiplet naturally
gains the dominant F-term, although the axion itself will not be light  since
 $\Lambda_h~>~\Lambda_{QCD}$ and cannot act as the QCD axion.

Hence, while we do not select a unique possibility, the wide range of UV
completions motivates our detailed study of the IR effects of SUSY
breaking through an axion F-term.  In particular, we consider the
extent to which it is possible to realize non-minimal patterns of soft
terms such as natural supersymmetry, non-universal gaugino masses, and
split supersymmetry from the effective interactions of such a
multiplet.  A single axion multiplet coupled to the SM captures these
effects, so in the spirit of minimality we will assume from now on
that just one axion multiplet gains an F-term and its couplings are
the dominant source of SUSY breaking in the MSSM.  It should be kept
in mind, however, that more than one axion multiplet may play a role
in SUSY breaking and its mediation.

\section{Effective Field Theory of Supersymmetric Axions}
\label{sec:couplings}

As previously discussed, in field theory constructions an axion arises as a
pseudo-Nambu-Goldstone boson from a spontaneously broken anomalous
global U(1) symmetry (the Peccei-Quinn symmetry), while in string models
axions can appear as fundamental moduli fields which automatically
have approximate shift symmetries. However, the origin of the axion is not
important from an effective field theory perspective: we simply
consider an axion to be a chiral multiplet, $A$, which respects a
shift symmetry at the perturbative level and may acquire a small mass
via non-perturbative effects.  The axion superfield $A$ contains the
pseudoscalar axion $a$ as well as the saxion $s$, axino $\psi_A$, and
auxiliary component $F_A$,
\begin{equation}
A= \frac{(s+ia) }{\sqrt{2}}+ \sqrt{2}\theta\psi_A  + \theta^2 F_A .
\label{eq:axionfield}
\end{equation}
At low energies the global PQ symmetry is realized non-linearly as a shift symmetry 
\begin{align}
A &\mapsto A+ i \alpha f,
\label{eq:trans}
\end{align}
where $f$ is the axion decay constant.  The low energy theory may also
contain matter charged under PQ; chiral multiplets $\Phi_i$ with PQ charge
$x_i$ transform as
\begin{align}
\Phi_i &\mapsto e^{i x_i \alpha} \Phi_i.
\label{eq:trans2}
\end{align}

The general low energy effective theory for an axion multiplet has
been discussed in \cite{Bae:2011jb}. We summarize the results here for
completeness and to establish notation.  The effective interactions in
a supersymmetric theory can be written as a sum of the gauge-kinetic
interaction $\Delta \Lag_{G} $, superpotential interactions $\Delta
\Lag_{S} $, and \kahler potential interactions $\Delta \Lag_{K}$ in
the Wilsonian effective action,
\begin{equation}
  \label{eq:17}
\Lag_{int} = \Delta \Lag_{G} + \Delta \Lag_{S}+ \Delta \Lag_{K}.
\end{equation}
The gauge-kinetic coupling
\begin{equation}
 \Delta \Lag_G = - \sum_n\int d^2\theta
\frac{C_{n}}{32\pi^2} \frac{A}{f} \tr\left( W^n_\alpha W^{n\alpha}
\right) +\rm{h.c.}
\label{eq:anomalouscoupling}
\end{equation}
defines the standard anomalous coupling to the MSSM gauge field
strengths $W_{\alpha}^n$, where $n$ labels the gauge group
$G_n$.\footnote{We use a gauge kinetic normalisation $\int d^2\theta \frac{1}{4 g^2}W_\alpha W^\alpha$.} 
While the term \eqref{eq:anomalouscoupling} is frequently considered
to give rise to the leading interactions of the axion multiplet, there
are additional super-potential and \kahler potential
interactions in the presence of chiral fields $\Phi$ transforming
under PQ. These couplings are {\em not} sub-leading and are in fact
crucial for the invariance of physical observables under field
redefinitions of the chiral multiplets.

Specifically, the renormalizable holomorphic superpotential couplings
are given by
\begin{equation}
\Delta \Lag_S = \sum_{ijk}\int d^2\theta\left( m_{ij} e^{-(x_{i}+x_{j})A/f}\Phi_i\Phi_j 
+ \la_{ijk} e^{-(x_{i}+x_{j}+x_{k})A/f}\Phi_i\Phi_j\Phi_k \right) + \mathrm{h.c.},
\label{eq:superpotentialcoupling}
\end{equation}
while the \kahler couplings to leading order in $1/f$ are\footnote{We work with a field
  expansion such that the saxion has zero VEV, in contrast to much of
  the string literature which is concerned with moduli stabilization
  where the minimum of the modulus (i.e. saxion/axion) potential is not
  known initially. This explains the difference between
  our form of the \kahler potential and that commonly given in string
  models.} 
\begin{equation}
\Delta \Lag_K= \sum_{i}\int d^4\theta\left(1+y_i \frac{(A+A^\dagger)}{f}  +z_i\frac{(A+A^\dagger)^2}{f^2}  \right) \Phi_i^\dagger \Phi_i.
\label{eq:kahlercouplings}
\end{equation}
The fields $\Phi_i$ may be light MSSM fields or additional
heavy vector-like pairs, with supersymmetric masses $m_{ij}$ and
trilinear Yukawa couplings $\la_{ijk}$. The axion multiplet can have further interactions with one
or more hidden sectors, and possibly a (small) mass arising from, for
example, string instanton effects, but we will not need to be explicit
about their details here.

There is a large number of parameters in
\eqref{eq:anomalouscoupling}-\eqref{eq:kahlercouplings}:
superpotential couplings $x_i$, \kahler couplings $y_i, z_i$, and
anomalous couplings $C_n$, which together specify all axion
supermultiplet couplings up to order $f^{-2}$.  However, these
parameters are not independent: under a chiral rotation
\begin{align}
\Phi_i & \mapsto e^{k_i A/f} \Phi_i,
\label{eq:redef}
\end{align}
they transform as
\begin{align}
  x_i &\mapsto x_i - k_i\nonumber\\
  y_i &\mapsto y_i +k_i \nonumber\\
  z_i &\mapsto z_i +y_i k_i + \frac{k^2_i}{2} \nonumber\\
  C_n &\mapsto C_n - 2\sum_i k_i T_n^{\Phi_i},
\label{eq:equivclass}
\end{align}
where $T_n^{\Phi_i}$ is the Dynkin index of the chiral fields $\Phi_i$ under the group $n$. The
transformation of $C_n$ is due to the Konishi anomaly
\cite{Konishi:1983hf}.  After such a redefinition the PQ symmetry is
still realized by \eqref{eq:trans} but with the new values of charges
$x_i$ as given in \eqref{eq:equivclass}. In particular, there is a set
of combinations invariant under field redefinitions,
\begin{align}
x_i &+y_i \nonumber\\
y_i^2 &- 2z_i\nonumber\\
C_n &- 2\sum_i x_i T^{\Phi_i}_n\nonumber\\
C_n &+ 2\sum_i y_i T^{\Phi_i}_n .
\label{eq:invariant}
\end{align}
For a theory with $N$ chiral superfields $2N+3$ of these combinations are linearly independent; as we will see, soft SM
sector masses will be proportional to these combinations when the
axion sector acquires an F-term. A UV theory determines a particular
set of couplings $x_i, y_i,z_i,C_n$, often in terms of a much smaller number of charges or couplings,
and the equivalence class of theories defined by the transformations \eqref{eq:equivclass} must
lead to identical expressions for physical observables. The choice of
basis can be made on grounds of clarity and convenience.

In anticipation of our study of the phenomenology of particular
models, it is interesting to consider the values such parameters may
typically take. Firstly the $x_i$ occur as a result of the charges of
fields under some group. Therefore these are expected to be either
natural numbers or exactly zero, with exponential separations or
non-integer values unlikely. They may be universal between different
generations, depending on the underlying theory (for example, as
discussed later due to brane localisation of matter), and additionally
may be universal within a generation, or vary for different
fields. For example in a GUT compatible model, within a generation the
fields $Q_L$, $u^c$ and $e^c$ are expected to have equal $x$, which
may be different to the value of $x$ for $d^c$ and $L$.

In our convention the parameters $C_n$ are order one when generated at
loop level in either field or string theory models.  Like the
coefficients $x$, these depend on the charges of fields under
symmetries, hence are typically natural numbers.  An alternative
scenario occurs in some string models with a tree level coupling
between the modulus multiplet and gauge fields, in which case $C_n$ is
$\mathcal{O}(32\pi^2)$.

Finally, the parameters $y$ and $z$ have a rather different physical
interpretation. These are not PQ symmetry breaking, and not
necessarily linked to charges under some group. In string models they
may be related to the modular weights of the fields, in which case
they are also typically natural numbers; however in the field theory
case they may be generated by integrating out additional
matter. Depending on the generation mechanism they may be universal
between generations (for example if they occur through integrating out
gauge fields), or alternatively may be able to vary over a very large
range. For example if the wavefunctions of different generations are
localized differently in a warped dimension, there could an
exponential variation in the magnitude of these
parameters. Alternatively if they occur by integrating out matter of
mass of the Planck mass they could be of magnitude $f/M_{\rm pl} \sim
1/100$ in the string axion case.

To clarify the meaning of a physical observable in the context of these basis dependent parameters, 
consider the divergence of the current associated to the PQ symmetry:
\begin{equation}
\partial_\mu J^\mu_{PQ} = \frac{g^2}{16 \pi^2}C_{PQn} F^{a\mu\nu}\tilde{F}^a_{\mu\nu} .
\end{equation}
Unlike the Wilsonian couplings $C_n$ which are changed by the field
redefinition \eqref{eq:redef}, the current divergence and therefor the
anomaly coefficient $C_{PQn}$ are physical quantities and must be
basis-independent. That this is so can be checked by computing the
divergence of the PQ current given by the sum of the Wilsonian
coupling $C_n$ and the anomalous diagrams
containing the chiral fields $\Phi$,
\begin{equation}
C_{PQn}= -C_n + 2\sum_i x_i T_n^{\Phi_i},
\end{equation}
which is one of the invariant combinations of \eqref{eq:invariant} as
expected. To say it another way, a chiral rotation of the fields
$\Phi$ changes the coupling $C_n$ in the Lagrangian but the change is
compensated by the shift in the couplings $x_i$ which generate an
anomaly through the well-known triangle diagrams.

Another example is the 1PI coefficient of the anomalous gauge-kinetic
coupling, which determines for instance the axino-gaugino-gauge
amplitude at leading order. Recall that by definition the 1PI action
includes, at fixed loop order, effects from integrating out all
momentum scales; any symmetries of the underlying theory must be
manifest in the 1PI generating functional. For the effective
interactions of the axion multiplet the 1PI coefficient at one loop is
given by \cite{Bae:2011jb}
\begin{equation}
C_{1PIn}\left(p\right)= -C_n- 2\sum_{m^2_i<p^2} y_i T_n^{\Phi_i}+ 2\sum_{m^2_i>p^2} x_i T_n^{\Phi_i}.
\end{equation}
The details of the calculation are not important for our work, but
we note the answer is indeed invariant under the transformations
\eqref{eq:equivclass}. This occurs for a very similar reason to the
invariance of $C_{PQn}$: Field rotations lead to changes in the
Lagrangian coupling $C_n$ which are cancelled by loop diagrams
containing the matter fields with charges $x_i$. It is interesting to
note that $C_{1PIn}$ is a function of momentum and the behavior
depends critically on the masses of the particles concerned; we will
encounter this mass dependance again in the SUSY-breaking gaugino
masses in Section~\ref{sec:gauginos}.

\section{Visible Sector Soft Terms}
\label{sec:F-term}

In Sections \ref{sec:uv} and \ref{sec:couplings} we have discussed the
motivation for axion mediation from the top-down perspective and
established an effective theory of a supersymmetric axion with an
emphasis on basis-independent physical observables. Now we turn to an
analysis of the soft terms induced by an axion multiplet that
participates in SUSY breaking dynamics and develops an F-term
expectation value $F_A$. We calculate the sfermion and gaugino
masses explicitly; many of our results can also be understood in an
elegant way from analytic continuation. Ultimately we find the soft
terms obtained can interpolate between traditional mediation
mechanisms such as gauge mediation and less explored possibilities such
as split supersymmetry.

\subsection{Sfermions}
Chiral matter fields feel SUSY breaking from the axion sector due to
\kahler and superpotential couplings. Depending on the supersymmetric
mass of the multiplet, we find the couplings $x$ in the superpotential
\eqref{eq:superpotentialcoupling} and $y,z$ in the \kahler
\eqref{eq:kahlercouplings} lead to sfermion masses proportional to
one of two invariant combinations: $x+y$ or $y^2 - 2z$. For a single
pair of fields $\Phi_{1},\Phi_{2}$ with a supersymmetric mass term
$m\Phi_{1}\Phi_{2}$, the mass matrix for the scalar components is given by
\begin{equation}
\mathcal{L}\supset-\left(\begin{array}{cc}
\phi_{1}^{\dagger} & \phi_{2}^{\dagger}\end{array}\right)\left(\begin{array}{cc}
m^{2}+\left(y_{1}^{2}-2z_{1}\right)\frac{F_{A}^{2}}{f^{2}} & \left(x_{1}+x_{2}+y_{1}+y_{2}\right)m\frac{F_{A}}{f}\\
\left(x_{1}+x_{2}+y_{1}+y_{2}\right)m\frac{F_{A}}{f} & m^{2}+\left(y_{2}^{2}-2z_{2}\right)\frac{F_{A}^{2}}{f^{2}}
\end{array}\right)\left(\begin{array}{c}
\phi_{1}\\
\phi_{2}
\end{array}\right) \label{eq:scalarmass},
\end{equation}
and associated masses of the scalar mass eigenstates 
\begin{equation}
\begin{aligned}
m_s^2= & m^{2}+\frac{1}{2}\left(y_{1}^{2}-2z_{1}+y_2^2-2z_2\right)\frac{F_{A}^{2}}{f^{2}} 
\\ &\pm \frac{F_A}{f}\sqrt{\left(x_{1}+x_{2}+y_{1}+y_{2}\right)m^2+\frac{F_A^2}{4f^2}\left((y_1^2-2z_1)-(y_2^2-2z_2)\right)^2}.
\end{aligned}
\end{equation}
where we denote the complex scalar component of $\Phi_i$ as $\phi_i$ and the
$\theta^2$ component as $F_i$. Clearly these masses are invariant
under the basis redefinition of \eqref{eq:equivclass}, as required.

There are two limits of relevance to our discussion. If the matter fields are heavy, $i.e.~m~\gg~({F_A}/{f})$, the
masses of the scalars are given by
\begin{equation}
m_s^2= m^2 \pm
m\frac{F_{A}}{f}\sqrt{\left(x_{1}+x_{2}+y_{1}+y_{2}\right)}.\label{eq:7}
\end{equation} 
An example are fermions charged under PQ that acquire a mass $M\lesssim\mathcal{O}(f)$ when PQ
is broken, such as the extra heavy matter in KSVZ axion
models \cite{Kim:1979if,Shifman:1979if}. If the heavy fields $\Phi_{Mi}$ with couplings to the
axion multiplet are also charged under the SM gauge group, they will
act as messengers in gauge mediation and give soft masses to sfermions
$\phi_{MSSM}$ at two loops,
\begin{equation}
\tilde{m}_{\phi,GM}^2= 2c_n \left(\frac{g_n^2}{16\pi^2}\right)^2 \sum_i
\Big(T^{\Phi_{Mi}}\left(x_1+x_2+y_1 +y_2\right)\Big)^2
  \left(\frac{F_A}{f}\right)^2 ,
\label{lightsquarks}
\end{equation}
where $c_n$ is the quadratic Casimir of $\phi$ under gauge group
$G_n$. These loop-suppressed contributions from messengers to soft
masses are important for MSSM fields which do not have a direct
coupling to the axion multiplet (i.e. the \kahler contribution
\eqref{eq:phikahler} $\tilde{m}^2_{\phi,K}=0$).

Alternatively, if the fields $\Phi_{1},\Phi_{2}$ are MSSM fields which satisfy $m\ll ({F_A}/{f})$
then the \kahler couplings, if present (for example as in the DFSZ case), will dominate leading to
masses
\begin{equation}
\tilde{m}^2_{\phi,K}=\left(y_{i}^{2}-2 z_{i}\right)({F_{A}^{2}}/{f^{2}}).
\label{eq:phikahler}
\end{equation}
These \kahler-mediated masses are in general one loop larger than the
gauge-mediated contributions \eqref{lightsquarks}. In the case of the
third generation quarks the superpotential terms proportional to $m^2$
may have a small but non-negligible effect. Such terms only appear
after electroweak symmetry breaking.

The superpotential couplings, if present, also induce trilinear terms,
\begin{equation}
\mathcal{L}\supset -\lambda_{ijk} \left(x_1+x_2+x_3+y_1+y_2+y_3\right) \frac{F_A}{f},
\phi_1 \phi_2 \phi_3
\label{aterms}
\end{equation}
which again depend on the combination $x+y$. These are of
particular importance in the case of the third generation sfermions
due to their large Yukawa couplings.

In explicit calculations, renormalization group evolution from the
SUSY breaking scale is important as in general we will take the
SUSY breaking scale to be relatively high. There are positive
contributions to scalar masses due to non-zero gaugino masses
\cite{Martin:1997ns} as well as negative contributions from other
scalar masses as in split family scenarios \cite{ArkaniHamed:1997ab}.

\subsection{Gauginos and gaugino screening}
\label{sec:gauginos}
Now we turn to gaugino masses, for which there are three significant
contributions: tree-level, \kahler-mediation, and
gauge-mediation. First, a coupling through the gauge kinetic term
$C_n$ leads to a tree level contribution once the axion supermultiplet acquires an
F-term\footnote{The mass includes a factor $4g^2$ due to our choice of gauge coupling
  normalisation and a factor of ${1}/{2}$ since
  this is a Majorana mass term.},
\begin{equation}
m_{C_n}= \frac{g^2 C_{n}}{16\pi^2} \frac{F_A}{f}.
\label{eq:gauginoanom}
\end{equation}

The anomaly coefficients $C_n$ can arise in the UV theory or from
anomalous diagrams with chiral matter charged under both PQ symmetry
and the gauge group $G_n$. In the case of the QCD axion the axion, of course, must couple to
QCD so $C_3 \neq 0$. Moreover in conventional 4D GUTs there is also the
condition $C_1=C_2=C_3$.
  
An additional mass contribution occurs due to the existence of a
counter-term in the gauge coupling that appears from any chiral
multiplet charged under the gauge group and which develops a non-zero
F-term. The presence of such a counter-term was first noted due to its
appearance in models of anomaly mediation
\cite{D'Eramo:2012qd,Cohen:2011aa,Dine:2007me,Bagger:1999rd}, and
leads to a mass contribution
\begin{align}
m_{c.t.}&= -\frac{g^2}{16\pi^2} \sum_i 2 T_n^{\Phi_i} \frac{F_i}{\left<\Phi_i\right>}
\label{eq:gauginocounterterm}
\end{align}
where $F_i$ is the F-term developed by a chiral multiplet $\Phi_i$
which also has a scalar expectation value
$\left<\Phi_i\right>$. Solving for the F-terms we obtain $F_i = y_i
\frac{F_A}{f} \left<\Phi_i\right> $, and therefore the gaugino mass
induced by this term is given by \footnote{There is a question of the
  validity of this term at the origin $\left<\Phi_i\right>=0$; the
  counter-term has been shown to persist in the same form in this case
  \cite{Cohen:2011aa}.}
\begin{align}
m_{c.t.}=-\frac{g^2}{16\pi^2} \sum_i 2 y_iT_n^{\Phi_i}
\frac{F_A}{f}.
\label{eq:gauginocounterterm2}
\end{align}
The two contributions \eqref{eq:gauginoanom} and
\eqref{eq:gauginocounterterm2} balance such that the gaugino mass is
unchanged under the anomalous transformations and demanding this is
one of the ways of deriving the expression for the counter-term
\cite{D'Eramo:2012qd}.

As noted in \cite{Dine:2007me}, at least part of what is known as
anomaly mediation is completely disconnected from any sort of
gravitational effect and appears in globally supersymmetric theories. The
contribution to the gaugino mass we see here is exactly the result of this
portion of anomaly mediation. While such a term is unimportant when considering the
classic SUSY spectra with gauginos of similar mass to sfermions, we will
later see that it can actually drive the creation of a split
SUSY spectrum.

Finally, there is a gauge mediated contribution from any chiral fields
which have direct couplings to the axion. Since the mass matrix for
the scalars is invariant under field redefinitions, so is the
contribution to the gaugino masses. The exact expression is
complicated, but we will mostly be interested in two extreme
regimes where the result simplifies. The first case is KSVZ-like \cite{Kim:1979if,Shifman:1979if} , 
where heavy states with $m \gg \frac{F_A}{f}$ couple to the axion; these act as messengers and give a
significant contribution to gaugino masses. The second case is
DFSZ-like \cite{Dine:1981rt,Zhitnitsky:1980tq}, in which the chiral fields with axion couplings are MSSM
fields. 

In the KSVZ-like case the masses are given by $m^2 \pm
m\frac{F_{A}}{f}\sqrt{\left(x_{1}+x_{2}+y_{1}+y_{2}\right)}$. Since
these fields form a vector-like pair,
$T_n^{\Phi_1}=T_n^{\Phi_2}=T_n^{\Phi}$; they lead to gauge
mediated contribution to the gaugino mass given by
\begin{equation}
m_{gauge}=-\frac{g^2}{16\pi^2} 2 T_n^{\Phi} \left(x_1+x_2+y_1+y_2\right)\frac{F_A}{f}.
\end{equation}
Adding this contribution to that obtained from the counter-term due to
the small F-term gained by this pair of fields we find a net mass
(where we are not including the mass contributions from any light
fields to be discussed shortly)
\begin{equation}
m_{1/2}= \frac{g^2}{16\pi^2}\Big(C_n-2  T_n^{\Phi} \left(x_1+x_2\right)  \Big)\frac{F_A}{f}.
\label{eq:KSVZgauginoresult}
\end{equation}
This is clearly redefinition invariant as expected; the extension to
more than one pair of states is straightforward. 

An interesting feature of \eqref{eq:KSVZgauginoresult} is that all dependence on
$y_i$ cancels. This is the  `gaugino screening' phenomenon 
\cite{ArkaniHamed:1998kj,Cohen:2011aa}, where \kahler couplings between
messenger fields and the SUSY breaking sector have no effect on
gaugino masses at leading order in $\frac{F}{m}$. 
This result can be understood from a holomorphic perspective. Gaugino
masses are given by the $\theta^2$ component of the real gauge
coupling $R$ which depends on the holomorphic gauge coupling $S$ as
\begin{equation}
R_n\left(\mu\right) = S_n + S_n^\dagger - \sum_l \frac{T_n^{\Phi_i}}{8\pi^2} \log\mathcal{Z}_i ,
\end{equation}
at a scale $\mu$. The holomorphic gauge coupling is given by
\begin{equation}
S_n\left(\mu\right)= S_n\left(\Lambda_{UV}\right) +
\frac{b_n}{16\pi^2}\log\left(\frac{\mu}{\Lambda_{UV}}\right) -\sum_i \frac{T_n^{\Phi_i}}{16\pi^2}\log\left(\frac{M_i}{\Lambda_{UV}}\right) ,
\end{equation}
where $M$ is the physical mass of the messengers which after analytic
continuation acts as a spurion, gaining a $\theta^2$
term. Additionally $M$ is the mass of the canonically normalized field
and therefore is given by $M=\frac{M_{s}}{\mathcal{Z}}$ where $M_s$ is
the mass that appears in the superpotential. Hence the real gauge
coupling is independent of $\mathcal{Z}$ and thus $y_i$ to leading
order.

The DFSZ-like case where the states are light is also
interesting. The gauge mediated contributions to gaugino masses are
completely negligible since for all MSSM fields $m\ll \frac{F_A}{f}$
\cite{Poppitz:1996xw}. Therefore the gaugino masses are given by sum
of \eqref{eq:gauginoanom} and the terms arising from the counter-term (where
the effects of heavy fields are now not included)
\begin{equation}
m_{1/2}=\frac{g^2}{16\pi^2}\left(C_n+2\sum_i y_iT_n^{\Phi_i}\right)\frac{F_A}{f} ,
\label{eq:DFSZgauginoresult}
\end{equation}
which now does depend on the $y_i$ but is still redefinition invariant
as required.  From an analytic continuation point of view it is clear
what has happened in this case too: the light states are not
integrated out and so
the real gauge coupling retains dependence on $\mathcal{Z}_i$ through
the term $R\supset -\frac{1}{8\pi^2}\sum_i T_n^{\Phi_i}
\log\mathcal{Z}_i$.

In a generic model with both heavy and light states the total gaugino mass is
\begin{equation}
m_{1/2}=\frac{g^2}{16\pi^2}\left(C_n-2\sum_{m_i\gg \frac{F_A}{f}}
 x_i T_n^{\Phi_i}+2\sum_{m_i\ll \frac{F_A}{f}}
  y_iT_n^{\Phi_i}\right)\frac{F_A}{f} ,
\label{eq:gauginomass}
\end{equation}
where the sum is over heavy states for the superpotential couplings
and \kahler couplings for light states, and the heavy states are
assumed to form vector-like pairs; this expression is also basis
independent for the same reason as the 1PI coefficient. Except for the
UV-sensitive coefficients $C_n$, the gaugino masses are proportional to
the gauge couplings squared as in  gauge mediation, although
particular values of the anomalous coefficients can lead to different
hierarchies. Compared to the sfermion masses, the gaugino masses can
be at the same scale (if there is no large \kahler contribution to the
scalars as in \eqref{lightsquarks} ) or a factor of $\frac{g^2}{16\pi^2}$ lighter
for scalars with mass determined as in \eqref{eq:phikahler}. Altering
the \kahler contribution can change the difference between gaugino and
scalar masses, but does not alter the ratios of gaugino masses
themselves, unlike models which interpolate between anomaly mediation
and gauge mediation (as for example \cite{Gupta:2012gu}).

\subsection{Higher order contributions}\label{highergauginos}

There is one more contribution to gaugino masses which, while not
phenomenologically important, is conceptually relevant. The real gauge
coupling actually includes a term due to an anomaly when the gauge
multiplet is rescaled to canonical normalisation \cite{ArkaniHamed:1998kj}
\begin{equation}
R\supset \frac{g^2}{8\pi^2}T^G_n\log\left(S+S^\dagger\right) .
\end{equation}
where $T^G_n$ is the Dynkin index of the adjoint representation of the gauge group $n$. To leading order, by inserting the leading dependence $S\supset \frac{1}{4g^2}-
\frac{C_n}{32\pi^2}\frac{A}{f}$, this gives a higher-order gaugino mass contribution
\begin{equation}
m_{1/2}'\supset -\frac{g^4}{8\pi^2} T^G_n
\frac{C_n}{16\pi^2}\frac{F_A}{f}.
\end{equation}
This is smaller than the tree level contribution by a factor of $\sim
\frac{g^2}{8\pi^2}$ therefore is not usually phenomenologically relevant, but superficially
appears not to respect redefinition invariance. The solution is found by including the next to leading order
gauge mediated contribution. This is given by \cite{ArkaniHamed:1998kj}
\begin{equation}
m_{1/2}'\supset -\frac{g^2}{8\pi^2}T^G_n \left(m_{gauge} + m_{c.t.} \right).
\end{equation}
Therefore this term is invariant under field redefinitions in the same
way as the leading order contribution. For heavy vector-like states
there is a cancellation between $m_{gauge}$ and $m_{c.t.}$ which gives
a higher-order, but still invariant, contribution to the mass,
\begin{equation}
m_{1/2}'\supset-\frac{g^4}{128\pi^4}T^G_n \left(C_n-2\sum_{m_i\gg \frac{F_A}{f}}
 x_i T_n^{\Phi_i}+2\sum_{m_i\ll \frac{F_A}{f}}
  y_iT_n^{\Phi_i}\right)\frac{F_A}{f} .
\end{equation}
Clearly, this depends on the effective field theory parameters in the
same combination as the leading contribution, $m_{1/2}$ given by
\eqref{eq:gauginomass}, hence can be written as
\begin{equation}
m_{1/2}' \supset -\frac{g^2}{8\pi^2} T_n^G m_{1/2} . \label{eq:gauginonext}
\end{equation}

There are additional corrections to gaugino and scalar masses from
gauge mediation at $\mathcal{O}(F^3/M^5)$, where $M$ is the messenger
mass. These do not appear in the expression from analytic continuation
which cannot capture higher order terms arising from super-covariant
derivatives. An expression calculated directly from loop diagrams can be
found in \cite{Poppitz:1996xw}, however for $f$ in the range of
interest these terms are normally negligible. In a region where these
corrections become important they typically increase the scalar masses
and decrease gaugino masses. One possibly interesting scenario is the
case where the leading contribution \eqref{eq:gauginomass} vanishes
identically, so the correction \eqref{eq:gauginonext} also vanishes;
then the dominant source of mass is the gauge mediated term of order
$(F^3/M^5)$, resulting in highly suppressed gaugino masses and a very
split spectrum.

\section{Notable Models}
\label{sec:spectra}

In the following subsections, we begin by exploring scenarios that
lead to unusual non-minimal SUSY spectra with possible relevance for
the LHC.  Axion mediation naturally interpolates between a
gauge-mediated type spectrum with gauginos and scalars at the same
scale (when the mediation is dominated by heavy fields charged under
PQ and SM fields) and a \kahler mediated split spectrum (when the
dominant axion couplings are to Standard Model fields). Specifically,
we consider the phenomenologically interesting case of split SUSY with
a sfermions a loop factor above the gauginos in mass (the so-called
mini-split scenario). In addition we look at `natural SUSY' models
with a mix of gauge and \kahler mediated contributions -- split
families or split gauginos, which ameliorate the tension between the
current LHC constraints and the requirements of naturalness.  We also
discuss some of the special features of SUSY-breaking mediation by the
QCD axion supermultiplet itself.

\subsection{Split Supersymmetry}
\label{sec:split}

\subsubsection{Scalars and gauginos}

With the continued null supersymmetry search results and the discovery
of a relatively heavy and SM-like Higgs, a possible interpretation of
current data is that the Higgs mass may be tuned, even in the presence
of supersymmetry.  Indeed, in the MSSM at least 10\% tuning is
inevitable given the high Higgs mass and stop and gluino limits from
the LHC \cite{Arvanitaki:2012ps}.  In Split Supersymmetry, the tuning
of the Higgs mass is one of the central ingredients; the rest of the
spectrum is minimal with scalar superpartners of the SM fermions
parametrically heavy, while gauginos and possibly higgsinos near the
TeV scale can preserve the success of gauge coupling unification and
provide a dark matter candidate
\cite{ArkaniHamed:2004fb,Giudice:2004tc,ArkaniHamed:2004yi}. In axion
mediation it is natural for the gauginos to be lighter than the
scalars by a loop factor or more, providing an attractive mini-split
spectrum, as we will now discuss.

In axion mediation, split supersymmetry is very straightforward to
achieve.  In the presence of \kahler couplings, $y$, between the axion
and MSSM chiral multiplets, sfermions gain masses $m_s^2\sim
\frac{y^2}{2} \frac{F_A^2}{f^2}$ while the gaugino masses are
suppressed by an extra loop factor $\frac{g_n^2}{16\pi^2}\sum_i
T_n^{\Phi_i} \sim 1/100$. This contribution to
gaugino masses appears as part of anomaly mediation, but is
independent of any supergravity effects, relying only upon the
presence of light fields charged under the SM gauge groups -- in this
case, the MSSM fermions themselves (this mechanism is dubbed \kahler
mediation in \cite{D'Eramo:2012qd}).

The sfermions in this case are relatively light---$10$-$1000\tev$---as
in the case of mini-split SUSY and in agreement with the measured
Higgs mass of $125\gev$ without the need for further model building to
raise the Higgs mass \cite{Arvanitaki:2012ps,Kane:2011kj}. In this mini-split case
it is either essential or advantageous that there is flavor structure
which prevents too-large flavor- and CP-violating observables arising
from these not very heavy sfermions of the first and second
generations. Importantly, in axion-mediation, this issue has a natural
solution as the axion can arise from \kahler moduli which couple
universally to scalars and for $f_a$ of string or GUT scale of
$10^{16}\gev$ or lower, these universal contributions dominate over
$M_{\rm pl}$-suppressed contributions. This feature alleviates the
flavor concerns that are present in generic anomaly mediation
scenarios
\cite{ArkaniHamed:2004fb,Arvanitaki:2012ps,Gupta:2012gu,ArkaniHamed:2012gw}. Note
that in many string constructions of supersymmetry breaking, e.g.,
KKLT \cite{Kachru:2003aw}, the non-universal complex structure moduli
are fixed by SUSY-preserving dynamics at high scale and, as explained
in the introduction, are thus are not usefully described as axion
multiplets. There is, however, a final light \kahler modulus which is
only stabilized by SUSY-breaking dynamics and which can play the role
of the axion supermultiplet of axion mediation.

\begin{figure}[t]
\begin{center}
    \subfigure[]{\includegraphics[trim = 0mm 40mm 0mm 0mm, clip, width
      = 0.45\textwidth]{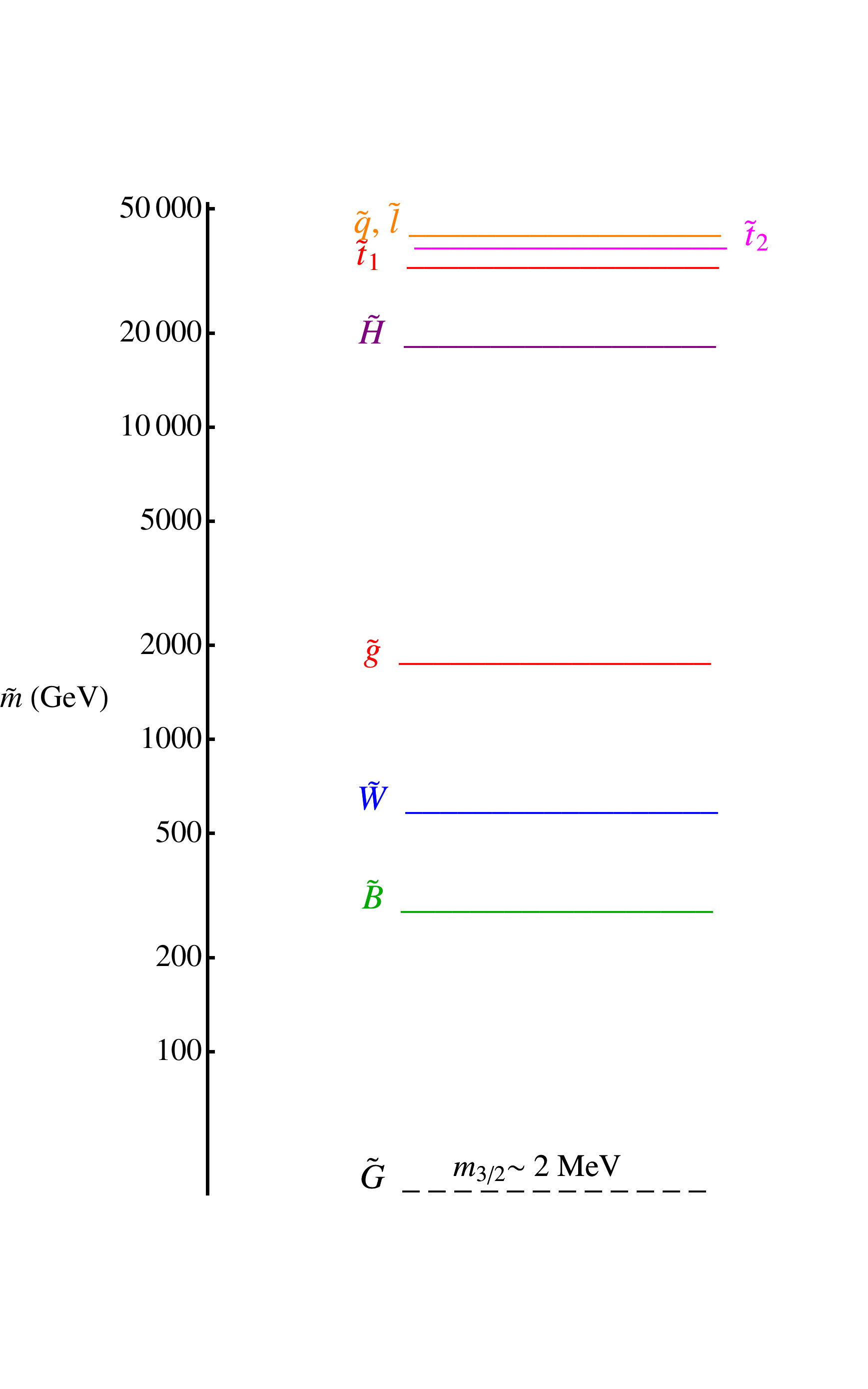}\label{fig:split1}}
\hspace{5mm}
    \subfigure[]{\includegraphics[trim = 0mm 40mm 0mm 0mm, clip, width
      = 0.45\textwidth]{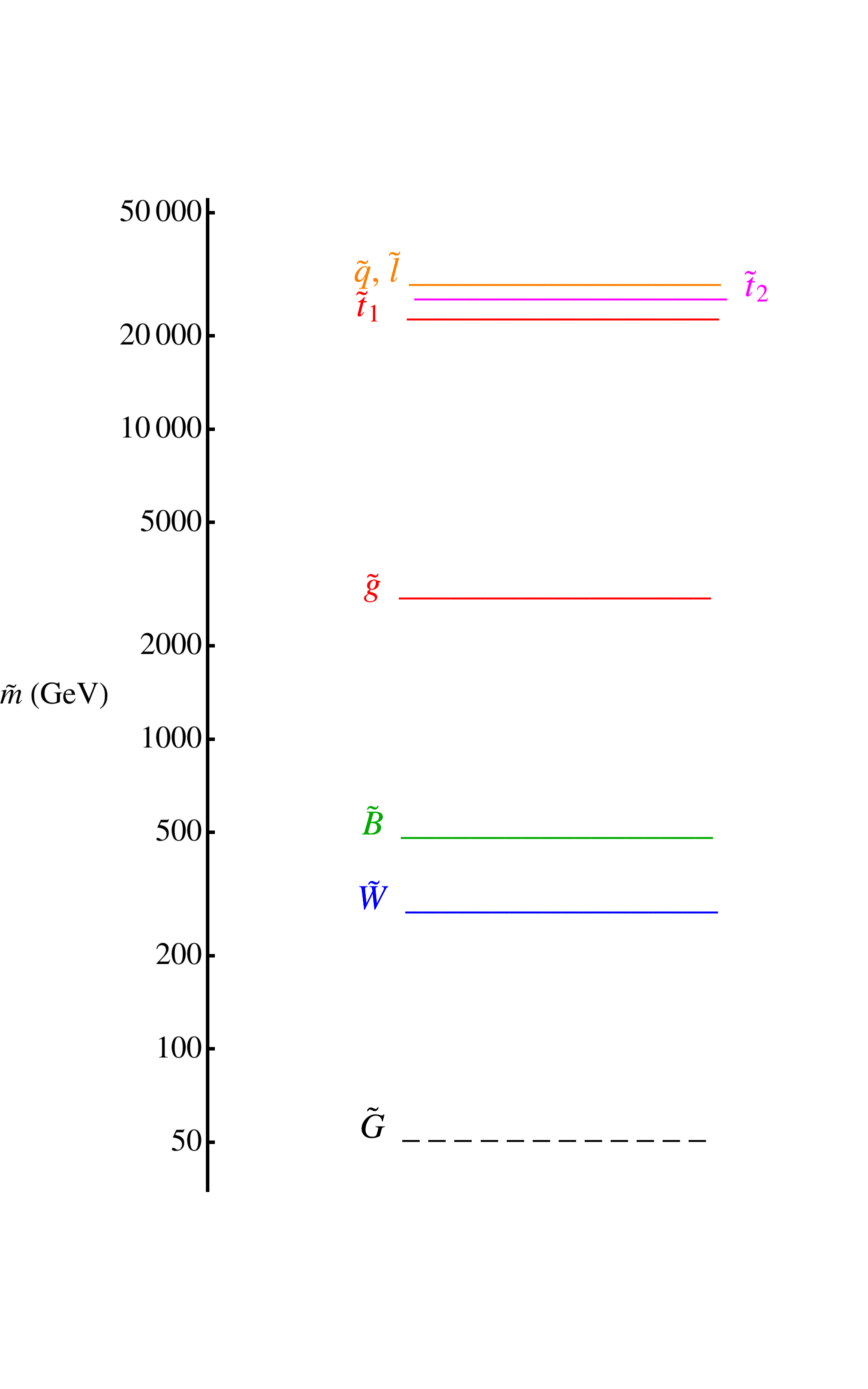}\label{fig:split2}}
    \caption{The spectra of MSSM soft masses, after running to the
      electroweak scale, in the case of \kahler mediation for (a)
      $f=10^{11}\gev$ in the `axion window' and (b) a `stringy' axion
      with $f=10^{16}\gev$. In the former case, $\mu$ is generated
      through PQ symmetry breaking, the gravitino is light, and the
      bino is the NLSP (assuming universal gaugino masses at the
      mediation scale).  The sfermions are at about $40\tev$, leading
      to a Higgs mass of $125\gev$ for $\tan\beta\sim 3$. In the
      latter stringy case, with similarly heavy scalars giving the
      observed Higgs mass, the gravitino has a mass of $\sim100\gev$,
      comparable to the gauginos, so either the gravitino or the
      gauginos could be the LSP.  Non-universal gaugino masses, which
      are a natural possibility in axion mediation, can lead to a wino
      NSLP avoiding potential conflict with late decay constraints in
      cosmology.} \label{fig:split}
   \end{center}
\end{figure}

Phenomenologically there are a range of possibilities depending on the
axion decay constant $f$ and the couplings in the \kahler
potential. First, consider an axion in the dark matter window
\cite{Preskill:1982cy, Dine:1982ah, Abbott:1982af}, $f \sim
10^{11}\gev$. Then to achieve TeV-scale gauginos with order one
couplings $y$, we take the scale of SUSY breaking to be $\sqrt{F} \sim
10^{8}\gev$, generating a typical spectrum as shown in
Fig.~\ref{fig:split1}. The scalars are at $40\tev$, leading to a
Higgs mass of $125\gev$ given $\tan\beta \sim
4$ \cite{Giudice:2011cg,Degrassi:2012ry}. All the gauginos could
potentially be discovered at the LHC. The relatively light scalars
mean that the gluino, if produced, would decay promptly inside the
detector and standard searches for the gluino apply
\cite{Gambino:2005eh}.  Of course, increasing $F$ or the \kahler
couplings increases the overall scale of the spectrum, taking the
gauginos out of experimental reach.

The gravitino in this case has a mass of $~2$~MeV; the gauginos are
stable on collider scales but will decay to the gravitino with a
lifetime of $\sim 0.3$~s, $i.e.$ sufficiently quickly not to pose a threat to
BBN \cite{Bailly2009,Jedamzik2006,Kawasaki2005,Kawasaki2008}.  The
gravitino will also likely be overproduced unless the reheat
temperature is very low, and depending on the exact value of $f$ the
gravitino and the axion will be co-dark matter.

The spectrum is changed for a stringy axion, $f \sim 10^{16}\gev$. For
TeV-scale gauginos and order one couplings $y$, a SUSY breaking scale
$\sqrt{F} \sim 10^{10}\gev$ gives a spectrum of the form shown in
Fig.~\ref{fig:split2}. The scalars are at $\sim30\tev$, with
$\tan\beta =5$.  The gravitino in this case has a mass of
$\sim100\gev$, comparable in mass to the gauginos, so either
the gravitino or the gauginos could be the LSP depending on
$\sqrt{F_A}$ and \kahler couplings $y$ and $z$.

If a neutralino is the NLSP in this scenario, it will decay to the
gravitino with a lifetime of $10^6$-$10^7$~s, in conflict with element
abundances in BBN as well as CMB measurement \cite{Feng:2003uy}. A
bino NLSP or LSP is thermally overproduced, leaving little room out of
these limits. However, for a stringy axion it is natural to have
non-universal gaugino masses as we will discuss in
Section~\ref{sec:splitgaugino}, and alternate neutralino NLSP admixtures
can alleviate the tension; for example, a $100\gev$ wino has thermal
abundance of less than $5\%$ of dark matter density, easing
cosmological constraints.

In the case that the gravitino is heavier than one or more of the
gauginos the gravitino lifetime is on the order of $10^7$-$10^8$~s,
also leading to tension with cosmology. Limits from BBN result in a
bound on reheating temperature of less than $10^6\gev$ unless the
gravitino is heavier than a TeV \cite{Kawasaki2005}. The latter is
disfavored for a spectrum with TeV-scale gauginos in
axion mediation as the axion scale would be $f>10^{17}\gev$.

Finally, for string values of $f\sim 10^{16}\gev$, the initial axion
misalignment angle has to be small not to over-produce dark matter,
with this environmental tuning becoming more severe as $f$
increases. Since there is no reason for the angle to be tuned further
than necessary, we again expect axion and neutralino or gravitino co-dark
matter.

\subsubsection{EWSB and Higgsinos}

Even though the Higgs mass is tuned, we must still consider the
details of the Higgs sector to ensure that electroweak symmetry
breaking takes place. There is a connection of PQ symmetry breaking to
the MSSM Higgs sector, which has a U(1) symmetry in the absence of the
$\mu$ term. If the MSSM Higgses are charged under the PQ symmetry
responsible for the axion, then breaking PQ will generate a $\mu$ term
for the Higgs, $\mu\sim\lambda {f^2}/M_{\rm pl}$, where $\lambda$ is
the coupling between the Higgses and the field which spontaneously
breaks PQ. Otherwise, if the Higgses are charged under a different
U(1) than the PQ symmetry of the axion, the $\mu$ term has to be
generated independently of the axion sector. 

In the `axion window', $f\sim 10^{11}\gev$, it is possible to identify
the PQ symmetry of the axion with that of the Higgses. For
concreteness, consider an example UV model
\cite{Bae:2011jb,Bae:2011iw}
\begin{equation}
  \label{eq:1}
  W = \lambda \frac{X^2}{M_{\rm pl}} H_uH_d + \lambda_s S (X Y - f^2).
\end{equation}
Here $X,Y$ are charged under PQ with charges of $-1$ and $+1$,
respectively; the two fields acquire vacuum expectation values of
order $f$. $H_u,\,H_d$ have PQ charges of $+1$ and other MSSM
particles have PQ charges set by the Yukawa interactions, e.g. $\{Q,L,
\overline{u},\overline{d},\overline {e}\}=\{-1,-1,0,0,0\}$. This
generates a $\mu $ term with $\mu \sim \lambda \frac{f^2}{M_{\rm pl}} $
when PQ is spontaneously broken.  Then
\begin{equation}
  \label{eq:15}
  \mu= \lambda \, (10^4 \gev)\left(\frac{f}{10^{11}\gev} \right)^2.
\end{equation}
For a dark matter window axion the $\mu $ term is at the same scale as
the sfermions. A large $\mu$ term also has an important effect on the
gaugino spectrum: threshold corrections proportional to $\mu$ change
the bino and wino masses by as much as $20\%$
\cite{Gherghetta:1999sw},
\begin{equation}
  \label{eq:5}
  \delta M_1 = \frac{3}{5} \frac{\alpha_1}{4\pi}\mu\frac{2m_A^2\sin
    2\beta}{m_A^2 - |\mu|^2}\log\left(\frac{m_A^2}{\mu^2}\right);
  \qquad \delta M_2 =  \frac{\alpha_2}{4\pi}\mu\frac{2m_A^2\sin
    2\beta}{m_A^2 - |\mu|^2}\log\left(\frac{m_A^2}{\mu^2}\right).
\end{equation}
We take this
effect into account in the spectrum in Fig.~\ref{fig:split}.

The low energy effective theory below the scale of PQ breaking can be
written as couplings in the \kahler and superpotential as in Section
\ref{sec:couplings}. Keeping the same Higgs charges, we also generate
$B_{\mu}$ through the term
\begin{equation}
  \label{eq:2}
  W \supset \mu \,e^{-(x_{Hu}+x_{Hd})A/f}H_uH_d + \mathrm{h.c.}
\end{equation}
so
\begin{align}
  \label{eq:3}
B_{\mu}&= \frac{\mu}{f}F_A (x_{Hu}+x_{Hd} ) = 2 \mu\frac{F_A }{f}.
\end{align}
Higgs fields coupled to the axion in the \kahler potential will give
masses to the scalar components of the same order as the sfermion
masses, $m_{Hu}^2\sim m_{Hu}^2 = y^2 (F_A/f)^2$. Then we have
$\mu\sim 10\tev$, $ m_{Hu}^2\sim m_{Hu}^2 \sim 40\tev$, which
results in successful electroweak symmetry breaking with
$\tan\beta\sim$~few. This mechanism also provides an upper bound on
sparticle masses: while all the masses in Fig.~\ref{fig:split1} can be
scaled up by increasing $f/F_{A}$ and take gauginos out of observable
reach, the Higgsinos are constrained to be below $100\tev$ by
gauge coupling unification \cite{Arvanitaki:2012ps}, so the connection with EWSB
ensures a relatively light spectrum.

Of course for the string axion, this mechanism would make $\mu$ far
too large unless the coupling $\lambda$ is quite small, $\lambda \sim
10^{-5}$.  Another possibility is to introduce heavy messengers which
couple to the Higgs to generate a $\mu$ term on the order of gaugino
masses. One example is coupling the Higgses directly to the messengers
(heavy fermions charged under PQ and SM gauge fields) as in models of lopsided
gauge mediation \cite{Dvali:1996cu, DeSimone:2011va}.  

\subsection{Split families and `Natural SUSY'}
\label{sec:nonuni}

With current limits on universal squark masses reaching close to 2
TeV, \cite{Chatrchyan:2011ek,:2012rz} an attractive scenario that maintains a natural solution to the hierarchy problem
is the split family class of models---also known as Natural Supersymmetry---in which the stops and
sbottoms are significantly lighter than the 1st and 2nd generation
squarks \cite{Dimopoulos:1995mi,Cohen:1996vb,Pomarol:1995xc}. Split
family models take advantage of the fact that light flavor
generations contribute more to the stringent limits due to high
production rates, while 3rd generation squarks contribute directly to
the tuning in the Higgs mass. Thus, separating the two issues by
raising the mass of the 1st and 2nd generations can relieve the strain
on naturalness from experimental limits (for some example models, see
\cite{Craig:2011yk,Craig:2012hc,Craig:2012yd,Gherghetta:2011wc}).

\begin{figure}[t]
\begin{center}
   \includegraphics[trim = 0mm 40mm 0mm 0mm, clip, width
      = 0.45\textwidth]{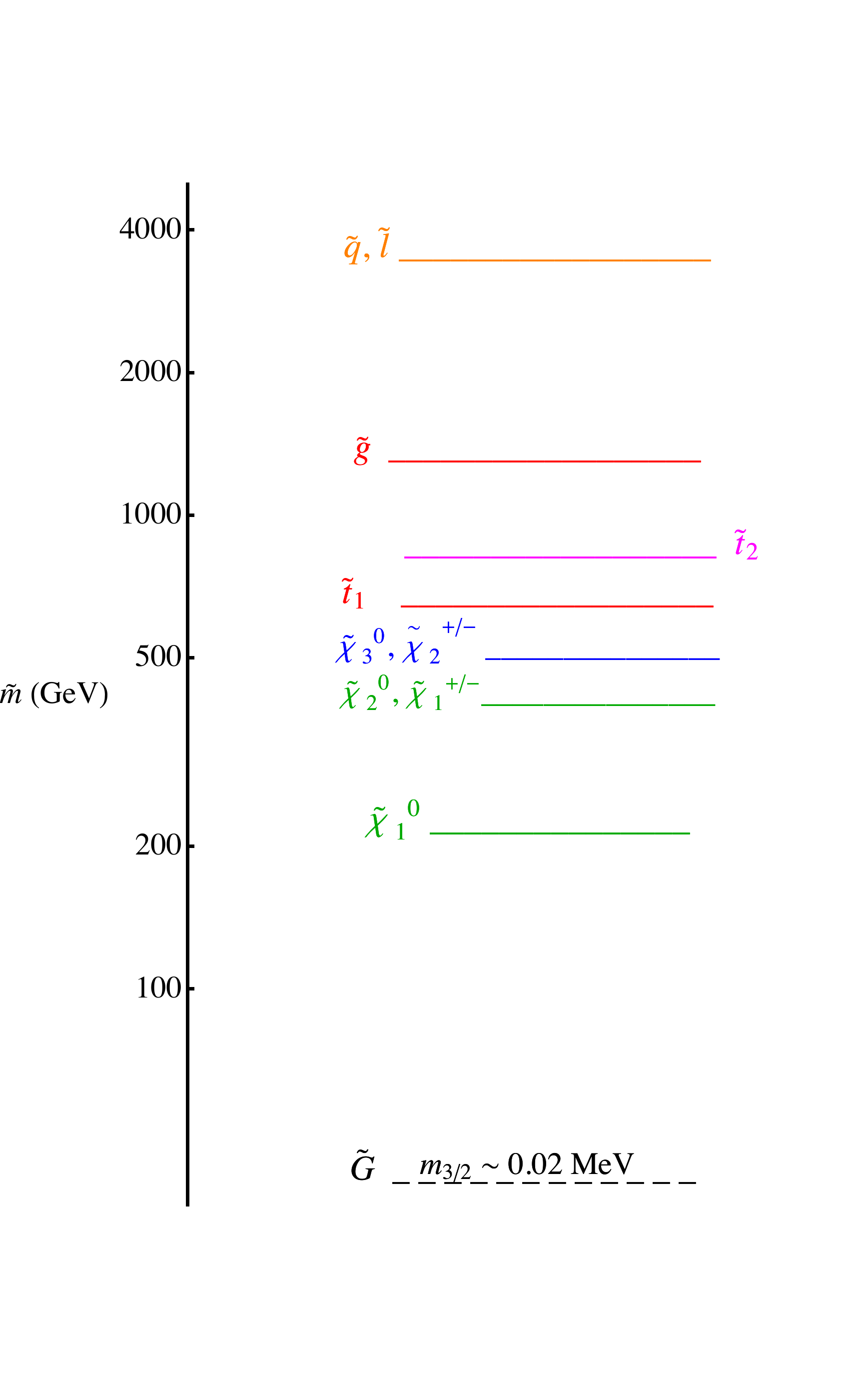}
      \caption{The low energy soft terms in the case of a split family
        spectrum, with $f_a\sim 10^{11}\gev$. TeV-scale visible
        sparticles require a SUSY breaking scale
        $\sqrt{F}\sim10^7\gev$. The first two generations have
        significant \kahler couplings to the axion multiplet, while
        those of the third generation are suppressed. Combined with a
        gauge mediated contribution this leads to mass splitting of
        the sfermion masses of about a factor of 5, taking the light
        generations out of experimental reach. Universal gaugino
        masses lead to a bino NLSP and gravitino LSP, while in the
        non-universal gaugino case different NLSPs are possible.
        Mildly heavy higgsinos lead to some tension with
        naturalness.} \label{fig:natural}
   \end{center}
\end{figure}

The most viable way of achieving a split family spectrum in axion
mediation is from an axion with couplings both to MSSM and new heavy
chiral multiplets. The first two generations are spatially separated
from the third along an extra dimension and gain a large mass from
\kahler couplings, while the third generation does not. Gauginos gain
mass both from the heavy multiplets and the \kahler couplings of the
first two generations. The third generation gains small masses through
gauge mediation from the heavy multiplets. A common issue with split
family scenarios is that heavy first and second generation sfermions
in combination with light gauginos tend to run third generation
scalars negative. The maximum splitting between generations is limited
to be a factor of $5$-$10$, so flavor problems arising from the first
two generation sfermions are not sufficiently solved
\cite{ArkaniHamed:1997ab}.  It is possible to address the flavor
problem in a relatively elegant fashion in which the first two
generations have an approximate symmetry protecting them against
too-strong flavor violations.  In particular, in the field theory case
\kahler couplings arise from integrating out additional heavy fields
which couple to the MSSM via new gauge interactions. This makes the
coefficients discrete parameters based on charge assignments that can
naturally be the same for some or all generations
\cite{flavouredsusy}.  Similar scenarios can also occur in the
string case.  Here we content ourselves with the overall gross
features of the resulting low-energy superpartner spectrum, a
representative example of which is shown in Fig.~\ref{fig:natural}; we
use \texttt{SOFTSUSY} to compute the low-energy spectrum
\cite{Allanach:2001kg}.  We leave the discussion of a complete model
for future work \cite{flavouredsusy} .

There are some drawbacks for natural SUSY in axion mediation due to
the high scale of SUSY breaking. First, in light of LHC limits,
a successful split families model has light stops and a gluino above
experimental limits ($m_{\tilde g} \geq 1.25\tev$ assuming a split
family spectrum,
e.g. \cite{ATLAS-CONF-2012-145,Aad:2012pq,CMS-PAS-SUS-12-026}); so in
raising the gluino mass we do not want to raise the stop
masses. Unfortunately the stop mass grows significantly through RG
running from a high scale. With even a somewhat low axion scale $f =
10^{11}\gev$, two-loop contributions from the gluino increase the
Higgs mass tuning by pulling up the stop mass such that
$m_{\tilde{t}}\sim (2/3) m_{\tilde{g}}$, regardless of the stop soft
masses at the SUSY breaking scale. This is just the general fact that
theories with a low cutoff scale are better for naturalness given
experimental constraints. In addition, the constraints on models with
a light gravitino and bino NLSP include charginos heavier than
$450\gev$ and stops above $580\gev$ \cite{Barnard:2012au} \footnote{We
  thank T. Gherghetta for calling our attention to these limits.}. Another
possible concern for the axion decay constant in the string motivated
window $f\sim 10^{16}\GeV$ is that additional generic
Planck-suppressed gravity mediated contributions to soft masses may be
large enough to cause tension with flavor constraints, however
understanding the extent to which this may be a problem requires a
full string construction. On the other hand, A-terms can naturally be
large in this case which can increase the Higgs mass within the MSSM.

\subsection{Split gaugino masses}
\label{sec:splitgaugino}

Typically, gauge or gravity mediation lead to gaugino masses falling
into a pattern proportional to their coupling constants $M_i \propto
g_i^2$. This leads to the well-studied case of fairly heavy
gluinos and (assuming R-parity conservation) substantial missing energy signals
from decays to the lightest neutralino of moderate mass.  In contrast in
axion mediation these patterns can be easily relaxed.

If the UV theory has an underlying GUT group and the PQ charges of
fields and axion multiplet couplings respect this group then the standard pattern of gaugino
masses is obtained. However, many models deviate strongly from this very
specific case. If, for example, the UV completion is an `orbifold-GUT' theory or a IIB type string model based on brane stacks
there is often no true 4D GUT symmetry, and matter which would normally be expected to
fall into a single irreducible SU(5) multiplet is now localized on different
branes  \cite{Hebecker:2001wq,Hall:2001pg}. Therefore these fields can very naturally have differing
charges under the PQ symmetry despite the fact that the success of supersymmetric
gauge-coupling unification is still naturally preserved and explained, at least in the orbifold-GUT case.
The anomaly coefficients may also naturally be non-universal
in an underlying heterotic string model \cite{Ludeling:2012cu}, or in other string constructions.

The phenomenology of non-universal gauginos can be interesting. Two
particular cases worth considering are $C_3 = 0$ and $C_1=0$. The
former case, which of course cannot be a QCD axion but is still
motivated in the axiverse picture of multiple axions, assumes an anomaly-free
SU(3).  The gluino has a
significant contribution to the fine tuning of the Higgs mass at two loops. 
With this in mind, certain mass ratios between $M_1,M_2$, and
$M_3$ have been studied and found to reduce the fine tuning of the
electroweak scale provided there is no fine-tuning in the UV theory to
set up these parameters \cite{Abe:2007kf,Horton:2009ed}. Not
surprisingly these mass ratios involve a gluino much lighter than
SU(2) gauginos at the SUSY breaking scale. In axion mediation, the
gluino can be much lighter than the wino and bino at the SUSY breaking
scale, provided $C_3=0$ and $x_i,y_i\ll C_{1,2}$, which can reduce the
fine tuning of the weak scale.

In the case $C_1=0$, the hypercharge U(1) is non-anomalous; then it
is possible to obtain a very light Bino, $M_1\ll M_2$ instead of the
usual relation $M_1 \sim \half M_2$. This loosens the indirect limits
on bino mass: without the correlation to chargino limits on $M_2$,
$M_1$ can be $50\gev$ or less. For high enough SUSY breaking scales
(such that the gravitino is heavier than the bino), the bino can then
be a viable dark matter candidate with the correct relic abundance, a
case that has been studied generally (see for example
\cite{Gabutti:1996qd,Bednyakov:1996ax,Belanger:2003wb,Dreiner:2009ic}).

\subsection{QCD axion mediation}
\label{sec:ax}

In a minimal model, the axion multiplet provides the solution to the
strong-CP problem as well as mediating supersymmetry breaking that is,
the pseudo-scalar component of the multiplet which acquires the
(dominant) F-term is the QCD axion.  As mentioned in the Introduction,
for the axion to solve the strong-CP problem its mass must arise
almost entirely from non-perturbative QCD breaking of the PQ symmetry,
and thus, given the constraints on the axion scale $f$, the QCD axion
mass must be small.\footnote{For completeness we note that there exist
  exceptions to this statement that, while unlikely in our view,
  cannot be definitively be excluded.  If, for example, the matter
  content of the UV theory were such that QCD becomes
  non-asymptotically free and strongly coupled {\em in the UV} then
  extra contributions to the axion mass could arise preserving the
  natural solution to the strong-CP problem.}  Such a light axion with
couplings to the SM is actively searched for, with the possibility of
its direct detection in laboratory experiments as well as indirect
detection via astrophysical observations
\cite{Beringer:1900zz,Kim:2008hd}.  As we will now discuss, in
addition to many aspects of the phenomenology of previous subsections,
this presents the exciting possibility of correlating axion detection
measurements to supersymmetric spectra; although of course making such
measurements experimentally, with sufficient precision to show
correlation, would be a very challenging task.

In more detail, one experimental observable is the anomalous axion-photon coupling,
\begin{equation}
  \label{eq:4}
  \frac{\alpha_1 C_{1}}{8\pi f} a F_{\mu\nu} \tilde{F}^{\mu\nu}.
\end{equation}
Searches for axions that have a two-photon vertex are especially
promising, including experiments which look for axions from the Sun
\cite{Dafni:2011eq} and the galaxy and place bounds of $C_{1}/f <
8.8\times 10^{−11} \gev^{-1}$ ($f_a > 3\times 10^8\gev$ for a QCD
axion). Laboratory searches include `shining light through a wall'
\cite{Raffelt:1987im} and microwave cavity experiments
\cite{Duffy:2006aa,Bradley:2003kg,Asztalos:2003px} which take
advantage of the axion to photon conversion in the presence of a
magnetic field and can place limits or potentially discover a light
axion.

Other detectable interactions are the derivative couplings of the axion to
fermions of the form $y_i (\partial_{\mu} a) \bar\psi \sigma^{\mu} \psi$, and 
$x_i a m_{\psi} \bar\psi \psi$ which arise in the SUSY context from
\kahler and superpotential couplings. From these couplings one derives
the basis-independent 1PI interaction between the axion and fermions,
\begin{equation}
-\left(x_1+x_2 +y_1+y_2\right)\frac{ m_{a}}{f} a \bar{\psi} \psi .
\label{eq:dcoupling}
\end{equation}

Interactions of the form \eqref{eq:dcoupling} with electrons and
quarks are experimentally relevant. The former can lead to excessive
white dwarf cooling which rules out $f<10^9\gev$, though there
may be a possibility that axion emission can improve fits of white
dwarf cooling models to the data \cite{Beringer:1900zz}.  If axions
couple to quarks, cooling by emission from nuclei can also be bounded
by constraints from SN 1987A to be $f>4\times 10^8\gev$.

From searches for superpartners at the LHC we can in turn learn about the
scale of supersymmetry breaking and the mass spectrum by measuring the
gravitino mass (in the case it is the LSP) and gaugino and sfermion
masses and correlate the mass spectrum with axion couplings. An extra
handle on the SUSY mediation is particularly appealing in split
supersymmetry, which has a short list of observables due to very heavy
scalars.  Gauginos masses will depend directly on the anomalous
coupling and the derivative coupling.  Then we can measure the
anomalous photon coupling $C_1$ and axion fermion couplings $y$ (and
continue to not observe flavor violations). Along with either the
gravitino mass or some knowledge from the Higgs sector this would give
enough information to prove mediation via this mechanism.

In the case of multiple axions the multiplet with the strongest
couplings to MSSM fields will be the one to dominate mediation when
all F-terms are comparable. So, it may be possible more generally to
discover the multiplet which dominates mediation in axion detection
experiments and provide evidence for axion mediation of supersymmetry
breaking.

\section{Axino as Goldstino and Cosmological Constraints}
\label{sec:goldstino}

In discussing the phenomenology of axion mediation in
Section~\ref{sec:spectra}, we focus on the minimal case in which a
single axion is the leading source of supersymmetry breaking; that is,
the axion multiplet is the only one in the theory that gains a
significant F-term. Then, the axino, as the fermionic component, is
also the goldstino in the theory; it is eaten by the gravitino through
the `super-Higgs' mechanism once we move to supergravity. The fact
that the axino is not an extra degree of freedom but part of the
gravitino multiplet is an additional benefit of SUSY breaking in the
axion sector, alleviating some tensions with cosmology as we discuss
below.

\subsection{The Axino and the Goldstino}

As is well known, couplings of the goldstino $\tilde{G}$ are fixed by
supercurrent conservation,
\begin{equation}
\mathcal{L} = i \tilde{G}^\dagger \bar{\sigma}^\mu \partial_\mu \tilde{G} - \frac{1}{F}\left(\tilde{G}\partial_\mu j^\mu + c.c. \right)
\end{equation}
where $j^\mu$ is the supercurrent of the other chiral and gauge
superfields \cite{Martin:1997ns},
\begin{equation}
j^{\mu} = \sigma^{\eta} \bar{\sigma}^{\mu} \psi_i \partial_{\eta}
\phi^{*i} - \frac{1}{2\sqrt{2}} \sigma^{\eta} \bar{\sigma}^{\rho}
\sigma^{\mu} \lambda^{\dagger} F_{\eta \rho} + \ldots. 
\end{equation}
The goldstino couplings determine the dominant interactions of the
longitudinal gravitino components: these are physical and are
important in collider phenomenology (if the gravitino is the LSP) as
well as in cosmology. However, these couplings do not readily appear
to be invariant under chiral rotations discussed in
Section~\ref{sec:couplings}, as they are fixed by the form of the
supercurrent and do not depend on whether the goldstino multiplet
transforms nonlinearly under a PQ symmetry.  However, the on shell
couplings depend on the superfields' masses, making the connection
between axino and goldstino interactions manifest.

This is clearest in the case of the axino-gaugino-gauge coupling; from
the goldstino Lagrangian,
\begin{equation}
\mathcal{L} \supset - \frac{i}{F}\frac{m^n_{1/2}}{2\sqrt{2}} \tilde{G}\sigma^{\eta} \bar{\sigma}^{\rho} F^n_{\eta \rho} 
\lambda_n^{\dagger}.
\end{equation}
By supersymmetry, the axino-gaugino-gauge is of the same form as the
gaugino masses \eqref{eq:gauginomass} as they both result from the
coupling $\int d^2\theta \frac{ \alpha_a}{16\pi }
\frac{C^a}{f}AW_\alpha W^\alpha$ along with loop contributions from
chiral fields. Then the goldstino coupling is manifestly invariant,
\begin{equation}
\mathcal{L} \supset -{i} \frac{\alpha_n}{8\sqrt{2}\pi f}\left(C_n-2\sum_h
  T_n^{\Phi_h}\left(x_{h1}+x_{h2}\right)+2\sum_l
  y_lT_n^{\Phi_l}\right) \tilde{G}\sigma^{\eta} \bar{\sigma}^{\rho} F^n_{\eta \rho} 
\lambda_n^{\dagger}
\end{equation}
 and since the axion has the only F-term, it is also the same in both descriptions.

For matter multiplets, couplings are between the scalar mass
eigenstates and the corresponding fermion mass eigenstates and
proportional to the mass difference between the two. To simplify the calculation we
consider one pair of vector-like fields. The fermions have
degenerate masses $m$ and the combination that couples to the scalar mass
eigenstate is the same combination of chiral multiplets. Since the
masses of the matter multiplets are again invariant under chiral
rotations, we find the coupling proportional to the masses,
\begin{equation}
\mathcal{L} \supset \frac{m}{f} \left(x_1+x_2+y_1+y_2 \right) \left(\psi_2 \phi_1+\psi_1\phi_2\right)+\frac{F}{f^2} \left((\psi_1\phi_1  \left(y_1^2-2v_1\right)+\psi_2 \phi_2 \left(y_2^2-2v_2\right)\right),
\end{equation}
which is of course redefinition invariant.  There are also
non-renormalizable couplings between two or more axinos and MSSM
fields; however, these are not constrained by the supercurrent
coupling and also are not phenomenologically relevant.

\subsection{Cosmology}

A frequent issue with an axion in the context of supersymmetry is the
cosmological axino problem; for a large class of models, the axino is
not protected by any symmetry, and upon SUSY breaking acquires a mass
of at least the gravitino mass \cite{Chun:1995hc,Cheung:2011mg}. If
the axion multiplet does not directly participate in SUSY breaking, it
will generically acquire a mass from supergravity effects in the
\kahler potential, 
\begin{equation}
  K \supset \int d^4\theta \frac{(A+A^\dagger)^2 (X+X^\dagger)}{M_{Pl}}
  \sim  \half m_{3/2}\, \psi_a\psi_a
    \label{eq:10}
\end{equation}
The mass is then at least of the order of the gravitino mass, unless
there are fortuitous cancellations or sequestering effects. Two weakly
interacting particles with comparable masses, the axino and the
gravitino, can be disastrous for cosmology: as pointed out in
\cite{Cheung:2011mg}, as at least one of them will be overproduced
over a large range of gravitino masses. In particular, the energy
density has opposite scaling with the mass for the axino and the gravitino.

In more detail, the energy density $m_{3/2}Y_{3/2}$ of the gravitinos
(where $Y_{3/2}$ is the gravitino abundance) is given by
\footnote{Here we assume that scattering dominates gravitino
  production for $m_{3/2} \gtrsim 10^{-4}\gev$, the medium- to
  high-scale range of supersymmetry breaking relevant for axion
  mediation.}
\begin{equation}
  \label{eq:12}
  m_{3/2}Y_{3/2} = m_{3/2} K M_{Pl}T_R \vev{\sigma_{3/2}
    v},
\end{equation}
where $ \vev{\sigma_{3/2} v}$ is the velocity-averaged scattering cross-section
and $K= \frac{n_s(T)}{n_s(T_R)} \frac{\sqrt{90}\zeta(3)}{\pi^3
  \sqrt{N_*}}$ is a numerical factor that depends on the change in
entropy density since reheating and the number of relativistic degrees
of freedom \cite{Moroi:1993mb}. The coupling strength of the gravitino
scales inversely with its mass, and the cross-section inversely with the mass squared: $\sigma_{3/2} \sim \frac{
  {\tilde m}^2}{m_{3/2} ^2 M_{\rm pl}^2}$. So the energy density is
inversely proportional to the gravitino mass,
\begin{equation}
  \label{eq:14}
    m_{3/2}Y_{3/2} \propto\frac{ T_R \tilde{m}^2}{m_{3/2} M_{\rm pl}}.
\end{equation}

The abundance is similar for the axino,
\begin{equation}
  \label{eq:13}
  m_{\psi a}Y_{\psi a} = m_{\psi a} K M_{Pl}T_R \vev{\sigma_{\psi a} v}
\end{equation}
However, for the axino the cross-section is independent of its mass and
depends only on $f$, $\sigma_{\psi a} \sim 1/f^2$.  For an axino with
the same mass as the gravitino, the energy density instead grows with its mass,
\begin{equation}
  \label{eq:14}
  m_{\psi a}Y_{\psi a} \propto \frac{m_{3/2} T_R M_{\rm pl}}{f^2},
\end{equation}
and the axino is overproduced at high $m_{3/2}$, while the gravitino
is overproduced at low $m_{3/2}$.

This results in an upper bound on low reheating temperature of $T_R <
3\times 10^5 \gev$ for all values of $m_{3/2}$ and an axion in the axion window, $10^9 \gev
< f < 10^{12} \gev$
\cite{Cheung:2011mg}. The bound is relaxed for higher values of $f$ in
the string axion regime, $f\sim 10^{16}\gev$: there, $T_R \lesssim
10^{9}\gev$ which is less restrictive but can still run into tensions
for instance with theories of high-scale baryogenesis.

The axino being eaten by the gravitino is very beneficial for
cosmology: as reviewed here, a light axino and gravitino together
result in low reheating temperature for much of the parameter space
causing tension between cosmology and supersymmetry breaking. Of
course, if the axino is the goldstino, the limit from a separate axino
abundance disappears, and high reheat temperatures are allowed for a
range of gravitino masses. Thus, an axion participating directly in
supersymmetry breaking dynamics and acquiring an F-term is a clear
and natural way to relax the friction between cosmology and a
supersymmetric axion.

\section{Discussion and Conclusions}
\label{sec:conclusion}

In this work we have considered the possibility that SUSY breaking is
mediated primarily through the interactions of a generalized axion
multiplet.  Non-minimal spectra, such as those of both natural
SUSY and split SUSY are simply realized.

We argued in Section~\ref{sec:uv} that from a UV perspective
axion-mediation of supersymmetry breaking is a very natural and
attractive possibility. This is especially (but not only) true in the
axiverse context, where we expect many light axion-like fields.
Following a thorough review in Section~\ref{sec:couplings} of axion
supermultiplet couplings in an effective theory language, we then
show in Section~\ref{sec:F-term} that the SUSY breaking
contributions to gaugino versus scalar masses depend on different
field-redefinition-invariant combinations of the axion supermultiplet
couplings. This leads to a straightforward implementation of split
supersymmetry, with heavy scalars separated from a loop factor from
and light gauginos, so realizing the attractive mini-split scenario,
as discussed in detail in Section~\ref{sec:split}.

A second interesting feature discussed in Section~\ref{sec:split} is
that the \kahler couplings which tend to dominate the visible sector
sfermion masses can be naturally universal, explaining the
non-observation of flavor violation due to supersymmetric partners.
In the string axion case this occurs when the axion is part of a
modulus which couples universally, as is the case in many explicit
constructions of SUSY-breaking in string theory, such as KKLT-like
scenarios with SUSY-breaking dynamics stabilizing the overall \kahler
modulus.  Moreover it is well known that to leading order \kahler
moduli know nothing about flavor structure and therefore couple
universally to all generations provided they are geometrically
localized in the same place \cite{Conlon:2006wz}.  Since the axion
must be part of a multiplet that is stabilized by non-SUSY-preserving
dynamics and, in addition, has a scale parametrically smaller
than $M_{\rm pl}$ (either $f\sim 10^{16}\gev$ in the string case, or
$\sim 10^9 - 10^{12} \gev$ in the `axion window') this
implies that the universal axion couplings naturally dominate the
non-universal gravitational and heavy moduli contributions to sfermion
masses.  In the field theory case, \kahler couplings
arise from integrating out additional heavy fields which couple to the
MSSM via new gauge interactions. This makes the coefficients discrete
parameters based on charge assignments that can naturally be the same
for some or all generations. On the other hand a natural SUSY
spectrum can be realized by localizing the third generation separately
from the first two, or by taking the discrete charge assignments to
differ across generations; some features of this scenario were
discussed in Section~\ref{sec:nonuni}.

In addition, if the axion couples to visible-sector fields in the
superpotential, as one would expect in the general case, then there
will be a characteristic pattern of (small) splittings among soft
terms. The sfermion mass squared will be split around a common central
value (either given by the \kahler couplings, or the gauge-mediated
contributions that occur in the KSVZ-like-axion case) by $2m_i
\frac{F}{f}$ where $m_i$ is the mass of the corresponding
fermion. Therefore lighter generations will be nearly degenerate and
the stops will be spread further. Observation of such a pattern would
give supporting evidence in favor of axion-mediated SUSY breaking.

Mediation of SUSY breaking via the QCD axion multiplet itself with
either axion scale in the axion window, or at the string value $f\sim
10^{16}\gev$, is experimentally interesting as well.  The axion itself
can be detected through its derivative couplings to matter or the
anomalous photon coupling \cite{Beringer:1900zz} and these couplings
can be correlated with the spectrum of superpartners to confirm some
features of axion mediation.
   
Axion mediation is also cosmologically beneficial in many cases as the
axino is eaten by the super-Higgs mechanism to give rise to the
massive gravitino.  This was studied in Section~\ref{sec:goldstino}
where we argue that cosmological constraints arising from axino and
gravitino overproduction are ameliorated.  There is an alternative
case of interest where, even though the axion multiplet dominates
mediation to the visible sector, there is another field with a larger
F-term. This may happen if the fields with the largest F-term do not
have strong couplings to the visible sector; then the axion is not the
goldstino and will not remain light. Even in this case there are
cosmological benefits: since the axion sector is now part of the
supersymmetry breaking sector the axino is expected to gain a mass of
order $\sim\sqrt{F}$, sufficiently large to be cosmologically
safe. This is similar to the cosmology studied in
\cite{Carpenter:2009sw}.

Finally, there are a number of possible extensions of this work. In
this paper we focus on the case of a single axion
supermultiplet which dominates the mediation of supersymmetry-breaking
to the visible sector.  However it is also possible that multiple
axion multiplets are involved in the mediation, and, moreover,
motivated by the same considerations that support the axiverse there
can be multiple SUSY breaking sectors, each with its own goldstino
\cite{Cheung:2010mc} or set of goldstinos \cite{Craig:2010yf}.
Related to this is the natural possibility that the axion under
consideration is the R-axion, a case that certainly deserves study.
We hope to return to these topics in a future work.

\acknowledgments{}

\noindent We are grateful to Asimina Arvanitaki, Savas Dimopoulos,
Tony Gherghetta, Xinlu Huang, James Unwin, and Giovanni Villadoro for
useful discussions.  This work was supported in part by ERC grant
BSMOXFORD no. 228169. JMR thanks the Stanford Institute for
Theoretical Physics for their hospitality during the early stages of
this work.  MB is supported in part by the NSF Graduate Research
Fellowship under Grant No. DGE-1147470 and acknowledges hospitality
from the Rudolf Peierls Centre for Theoretical Physics at Oxford where
parts of this work were completed.  The authors also thank the CERN
Theory Group for their hospitality.

\bibliography{AxionMediation2013_final_v2}
\bibliographystyle{JHEP}

\end{document}